\documentclass[a4paper,twoside]{article}

\usepackage[table]{xcolor}
\usepackage{epsfig}
\usepackage{subfigure}
\usepackage{calc}
\usepackage{amssymb}
\usepackage{amstext}
\usepackage{amsmath}
\usepackage{amsthm}
\usepackage{array}
\usepackage{multicol}
\usepackage{multirow}
\usepackage{pslatex}
\usepackage{apalike}
\usepackage{SCITEPRESS}
\usepackage[small]{caption}
\usepackage{supertabular}
\usepackage{rotating}

\begin{document}
\def\rot{\rotatebox}

\title{Leveraging video annotations in video-based
  e-learning}

\author{\authorname{Olivier Aubert, Yannick Pri\'e and Camila
    Canellas} \affiliation{University of Nantes, LINA - UMR 6241, Nantes, France}
  \email{olivier.aubert, yannick.prie, camila.canellas@univ-nantes.fr}
}

\keywords{E-learning, MOOCs, Video Annotation, Pedagogical Processes.}

\abstract{The e-learning community has been producing and using video
  content for a long time, and in the last years, the advent of MOOCs
  greatly relied on video recordings of teacher courses. Video
  annotations are information pieces that can be anchored in the
  temporality of the video so as to sustain various processes ranging
  from active reading to rich media editing. In this position paper we
  study how video annotations can be used in an e-learning
  context - especially MOOCs - from the triple point of view of
  pedagogical processes, current technical platforms functionalities,
  and current challenges. Our analysis is that there is still plenty
  of room for leveraging video annotations in MOOCs beyond simple
  active reading, namely live annotation, performance annotation and
  annotation for assignment; and that new developments are needed to
  accompany this evolution.}

\onecolumn \maketitle \normalsize \vfill

\section{\uppercase{Introduction}}
\label{sec:introduction}

While video material had been used for several decades as a learning
support, the development of web-based e-learning first caused a
setback in the usage of pedagogical videos, due to lack of network
bandwidth or standardized formats and software. However, video
streaming, video hosting and the dissemination of capture and editing
tools have come along and supported the exponential growth of video
usage on the Web. Again video became an important component of
e-learning setups, through the OpenCourseWare movement and the recent
advent of Massive Online Open Courses (MOOCs).

Video annotations (section~\ref{sec:video-annotations}) are
information pieces that can be anchored in the temporality of the
video so as to sustain various processes ranging from active reading
to rich media editing~(section~\ref{sec:systems}). Our main interest
in this position paper is related to how video annotations are and can
be used in e-learning context - especially MOOCs - from the triple
point of view of pedagogical processes~(section~\ref{sec:processes}),
current technical platforms functionalities~(section~\ref{sec:uses}),
and current challenges~(section~\ref{sec:challenges}). One of the most
important results of our analysis\footnote{This work has received a French government support
    granted to the COMIN Labs excellence laboratory and managed by the
    National Research Agency in the "Investing for the Futures"
    program ANR-JO-LABX-07-0J.} is that there is still plenty of
room for leveraging video annotations in MOOCs beyond simple active
reading, namely live annotation, performance annotation and annotation
for assignment; and that technological improvements are needed to
accompany this evolution.

\section{\uppercase{Video annotations}}
\label{sec:video-annotations}

Active reading is a process where a reader assimilates and re-uses the
object of his reading, as part of his knowledge
work~\cite{waller2003}.  The goals may be the exploration of a
document, its enrichment or its analysis, for oneself or within a
collaborative activity. Active reading usually relies on annotations,
that add some information to a specific section or fragment of the
target document, and can thereafter be reused along it for searching,
navigating, repurposing, etc.

The link between the annotation and the original document may be more
or less explicit, from the handwritten note in the margin of a book to
the note taken on a notebook while watching a movie (which will
involve more work from the annotator to specify the targeted
information). Two main components of an annotation are usually
considered: its content and its anchor. The content may take any form,
as long as the underlying support allows it, and can be more or less
structured. Anchoring will depend on the nature of the annotated
document, and will be more or less explicit and easy to navigate.

We specifically position ourselves in this article in the audiovisual
context. Video documents present a number of specific or more
stringent issues. First, contrary to text content, they do not have
any implicit semantics: any active reading must thus be mediated
through annotations that provide an explicit information layer along
the document. Second, contrary to text reading, the reading speed is
imposed by the player. The annotation process then requires some kind
of interruption or interference from the viewing process, and this
conflict between the temporality of the video and that of the
annotation process must be addressed somehow.

\begin{figure}[!h]
  \centering
  \includegraphics[width=\linewidth]{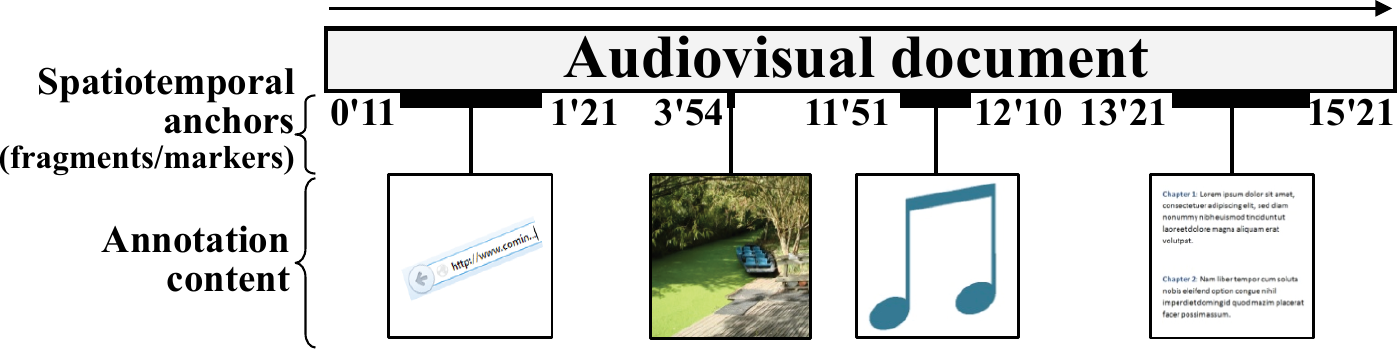}
  \caption{Video annotations anchors and content}
  \label{fig:video-annotation}
 \end{figure}

 A video annotation is composed of data explicitly associated to
 a spatiotemporal fragment of a video. As illustrated in Figure 1, the
 spatiotemporal anchors define at least a timestamp (in the case of
 duration-less fragments, specifying a single time in the document,
 e.g. 3’54), but more generally a begin and an end timecodes
 (e.g. 13’21 and 15’21). They may additionally address a specific
 static or dynamic zone of the displayed video (e.g. a rectangle shape
 that would follow a player in football video).

 Video annotation content data can be of any type. Textual data is
 most often used, since it is the easiest to produce and to consume,
 but any content (audio, images, video, key/values...) can as well be
 associated. For instance, in a language learning context, the tutor
 can take textual notes about a video recording of a session, and also
 annotate by recording some spoken words to indicate the correct
 pronunciation. Annotations can also be articulated through some
 structure, such as a type classification: a feature movie analysis
 could for instance define different annotation types like “Shot”,
 “Sequence” or “Character appearance” (see
 figure~\ref{fig:video-annotation}).

\section{\uppercase{Using video annotations}}
\label{sec:systems}

\begin{figure*}[htp]
  \centering
  \includegraphics[width=\linewidth]{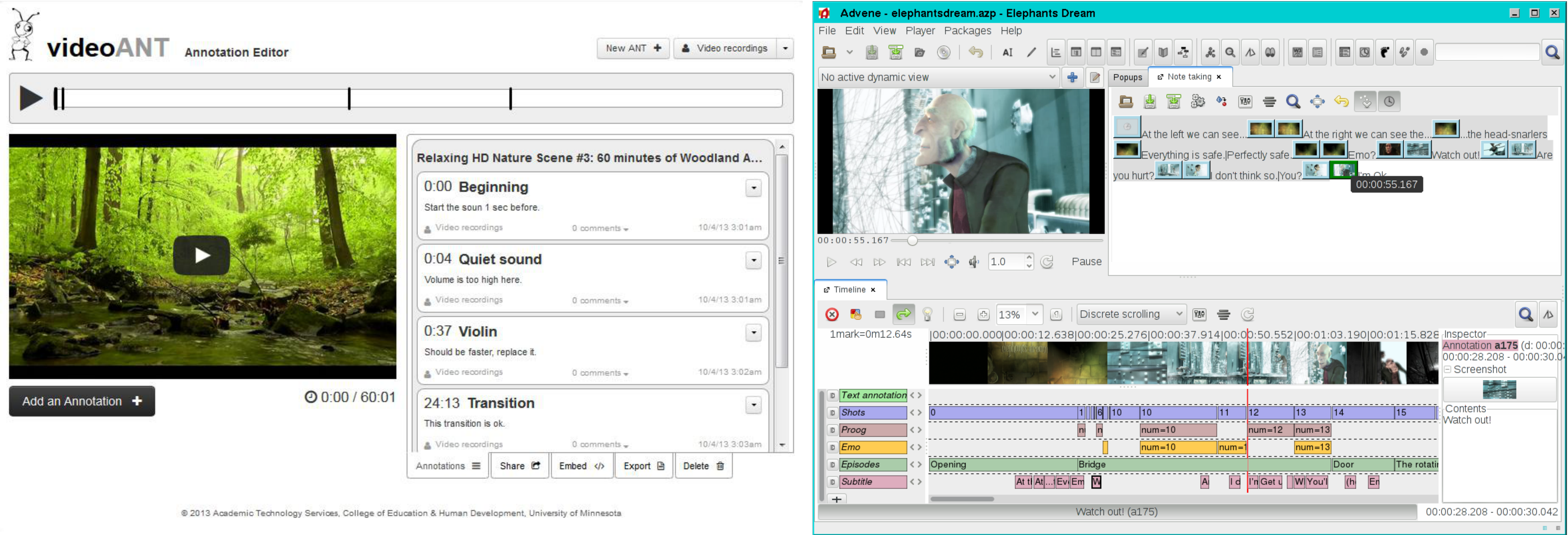}
  \caption{On the left, the VideoAnt video online annotation system
    displays the video with some annotations organised in a list. On
    the right, the Advene video annotation tool features several
    annotation display and creation interfaces: here a timeline at the
    bottom and a temporalized note-taking view on the right of the video.}
  \label{fig:outils}
 \end{figure*}

Annotations can be created manually or automatically. Manual creation
involves various user interfaces, depending on the nature of the task
and on the information that has to be captured (see VideoAnt and
Advene in figure~\ref{fig:outils}). Annotations may also automatically be created by
extracting features from an actual video document (through speech
recognition, or automatic shot detection for
example)~\cite{Nixon2013TVs-Future-is-L}, or by capturing
synchronized information during the very recording of the
document. This last case is used for instance when recording
information about the activity of a user: an ergonomics researcher
studying the use of a software can capture a recording of the user
screen while using the software, along with more discrete information
capture from the software (button clicks, file openings, etc.).

Beyond information retrieval and search, which is routinely carried
out in active reading activities or in video monitoring systems, video
annotations can be used in a variety of ways, such as enrichment and
document creation.

 Enrichment of the rendering of the video document is not new:
 subtitles can indeed be considered as video annotations, that are
 displayed as a caption over the video. But such overlays can also be
 graphical, to attract the attention of the viewer on a specific
 visual element. Video enrichments produced from the annotations can
 also be placed along the (original) video player, to produce a
 navigable table of contents for instance.  Video annotations can also
 be used to create other documents, be it other videos as it is the
 case in video summarization (automatic or guided); or more rich-media
 documents as an article illustrated with some fragments
 (through the annotations) of the video; or even an annotation-based
 hypervideo that permits the navigation from one video to the
 other. Eventually, collaborative activities can greatly benefit from
 annotations, that here serve as an interpretative layer between
 participants.

 These different types of uses can be put to use in different
 application domains.  \textbf{Video archives} (e.g. TV, surveillance)
 can propose an enhanced access to their collections through video
 annotations, allowing to find specific video fragments. The
 \mbox{\underline{Yovisto platform}}\footnote{Underlined terms have an
 associated URL given in the Webography annex at the end of the article.}~\cite{Waitelonis2012Towards-explora} offers for example
 access to video through semantic annotations, allowing to look for
 specific location, people, events... \textbf{Sports analysis} also
 greatly relies on video material, which can be used in a reflective
 way by offering the sportsman to view his own performance, or to
 analyse the behaviour of adversaries on recordings of previous
 competitions. Many applications such as
 \mbox{\underline{EliteSportsAnalysis}}
 or \mbox{\underline{MotionView Video Analysis Software}} offer tools to annotate and analyse sport
 performances.  \textbf{Research on activity} in domains such as
 ergonomics, animal behaviour, linguistics, etc. also uses annotation
 software, since researchers need to perform a precise analysis of
 video recordings. There exist a number of research tools such as
 \mbox{\underline{Advene}},
 \mbox{\underline{Anvil}} or
 \mbox{\underline{Transana}}, as well as
 commercial offers like
 \mbox{\underline{Noldus}}. They all offer
 annotation capabilities accompanied by various visualisation and
 analysis capabilities.  \textbf{Pedagogy} is obviously an important
 domain for video annotation practices. First, any matter dealing with
 videos content as learning material, such as language or movie
 courses \cite{Aubert2005Advene-active-,Puig2007Lignes-de-temps2}, can
 benefit from the usage of annotations on a course support, as in
 \mbox{\underline{VideoNot.es}}.  Second, discussion
 about self-reflective activities can be enhanced by annotation-based
 tools. For instance,
 \mbox{\underline{VideoTraces}}
 has been used for a long time in dance
 courses~\cite{Cherry2003Using-a-Digital,Bailey2009Dancing-on-the-}
 for annotating dance sessions. More generally, video recordings of
 learners presentations or interaction are used to implement
 self-reflection activities in
 classrooms~\cite{Rich2009Video-Annotatio}, supported by a number of
 tools such as
 \mbox{\underline{VideoTraces}},
 \mbox{\underline{CLAS}} or
 \mbox{\underline{MediaNotes}}.

 It appears that a great number of practices have been experimented in
 different contexts and application domains, from the 1990s VHS based
 reflective activities to the more recent collaborative and online
 video analyses. The experience accumulated on these tools and
 practices could fruitfully be incorporated in e-learning.

\section{\uppercase{E-learning processes based on video annotations}}
\label{sec:processes}

In order to assess in what measure we can leverage the existing
experience in video annotation systems in an e-learning context, we
organize these processes along four classes of
scenarios.

For \textbf{active reading with annotations}, video is a learning
material whose content has to be assimilated or evaluated. This task
is carried out through an iterative process, dedicated to the analysis
of the audiovisual source through its enrichment with annotations and
the definition of appropriate visualisations. In a learning context,
students may annotate the video material by taking notes, by
themselves or collaboratively. Conversely, teachers may use the same
techniques for evaluating and grading videos produced by students. And
both learners and teachers can engage into a discussion about a video
through annotations.

\textbf{Live annotation} occurs during a live lecture, which is
recorded and annotated at the same time. Students take notes during
the lecture, and reuse these notes as a basic indexing system when
replaying the recording. Teachers may also let students ask questions
through annotations, and answer them at the end of the
lecture~\cite{Betrancourt2011Assessing-the-u}.

\textbf{Performance annotation} also implies video as a trace of a
performance, be it recorded in a conventional classroom or during a
synchronous online session. The recording may already be augmented
automatically by the capture of annotations containing information
about the activity (sent documents, chat messages, etc). Based on this
recorded video and activity trace, students may annotate their own
performance, for self-reflection or for sharing an analysis with their
teacher~\cite{Rich2009Video-Annotatio}. Teachers may as well annotate
their own performance in a self-reflective way, to improve their
practice (ibid). Eventually, students may annotate a recorded course
for suggesting improvements or pointing out difficult sections. The
teacher can use that feedback when preparing the next course or the
next version of the same course~\cite{Sadallah2013A-Framework-for}.

Eventually, in \textbf{annotation for assignment}, the video is a
material that has to be used to prepare an assignment (a feature
movie, a recording of new for media analysis, etc). The work may
require students to analyse some aspects of the video, and produce
annotations reflecting their analysis. The annotations are then later
assessed by the teacher~\cite{Wong2010Teaching-develo}. Further, the
annotations resulting from the analysis may be reused to produce a new
document, like an abstract or a video collage. At Columbia University,
students use the MediaThread
platform~\cite{Bossewitch2011Teaching-and-Le} to produce critical
video composition or critical multimedia essays, by combining
annotations. The teachers then assess their productions.

We can identify distinguishing features between these different
classes of scenarios, considering on the one hand the status of the
video, and on the other hand the actors producing and using the
annotations. The annotated video can be a base learning material, such
as a movie or a documentary to study, or can be a recording of a
lecture (which the students may or may not have seen live). It can
also be the recording of student contributions.  The actors producing
and/or using the annotations can be the students, the teachers,
student colleagues or teacher colleagues, or even the general public.

\section{\uppercase{Annotation uses in e-learning systems}}
\label{sec:uses}

Table~\ref{table-annotations} provides an overview of the various functionalities related to
video annotation offered by mainstream MOOC platforms\footnote{Based
  on courses available in late December 2013. Some platforms such as Udemy
  were not considered due to the fact that they do not have an open
  access. This table will be actualized for the final
  version of the paper.  A more detailed version is available on the
  web and constantly actualized on
  http://comin-ocw.org/video-annotations/platform-features/.}
or more specifically dedicated tools.
\definecolor{light-gray}{gray}{0.95}
\begin{table*}
\scriptsize
\caption{Annotation-related functionalities offered by mainstream MOOC platforms or dedicated tools}
\label{table-annotations}
\begin{supertabular}{p{.5cm}|p{5cm}|*6{>{\centering\arraybackslash}p{.3cm}}|*6{>{\centering\arraybackslash}p{.3cm}}}

\multicolumn{1}{l}{~} & \multicolumn{1}{l}{~}
& \rot{90}{\textbf{\underline{EdX}}}
& \rot{90}{\textbf{\underline{Coursera}}}
& \rot{90}{\textbf{\underline{Canvas Network}}}
& \rot{90}{\textbf{\underline{Khan Academy}}}
& \rot{90}{\textbf{\underline{Iversity}}}
& \multicolumn{1}{l}{\rot{90}{\textbf{\underline{Open2Study}}}}
& \rot{90}{\textbf{\underline{VideoANT}}}
& \rot{90}{\textbf{\underline{VideoNot.es}}}
& \rot{90}{\textbf{\underline{Annotated HTML}}}
& \rot{90}{\textbf{\underline{Mediathread}}}
& \rot{90}{\textbf{\underline{YouTube}}}
& \rot{90}{\textbf{\underline{Matterhorn Player}}}
\\

\hline

\multirow{6}{*}{\begin{turn}{90}Visualization\hspace*{.2cm}\end{turn}}
 & \textbf{Multilanguage subtitles} & ~ & X\textsuperscript{1,2} & X & X\textsuperscript{1} & X & X &
~ & ~ & ~ & ~ & X & X\\

 & \textbf{Transcription} & X & X & X & X & ~ & X &
~ & ~ & ~ & ~ & X & ~ \\

 & \textbf{Other synchronized enrichments} \tiny{(e.g. slides)} & ~ & ~ & ~ & ~ & ~ & ~ &
~ & ~ & ~ & ~ & ~ & X\textsuperscript{3}\\

 & \textbf{List of annotations} (navigable) & ~ & ~ & ~ & ~ & ~ & ~ &
 X & X & X & ~ & ~ & X\\

 & \textbf{Timeline with annotations} & ~ & ~ & ~ & ~ & ~ & ~ &
X & ~ & ~ & ~ & ~ & X\\

 & \textbf{Interactive enrichments} \tiny{on the video or aside (e.g. embedded questions, alternative endings, hypertext links to external content)} & ~ & X\textsuperscript{4} & ~ & X\textsuperscript{4} & ~ & ~ &
~ & ~ & ~ & ~ & X\textsuperscript{5} & ~ \\

\hline \multirow{7}{*}{\begin{turn}{90}Editing/sharing features\hspace*{0.1cm}\end{turn}}
 & \textbf{Comment} \tiny{(about the whole video)} & X & ~ & ~ & X & X & ~ &
~ & ~ & ~ & X & X & X \\

 & \textbf{Video markers} \tiny{(single timecode + comment)} & X\textsuperscript{6} & ~ & ~ & X\textsuperscript{6} & ~ & X\textsuperscript{6} &
X &  X &  X & ~ & X & X \\

 & \textbf{Internal annotation tools} \tiny{(natively on the platform)} & ~ & ~ & ~ & ~ & ~ & ~ &
X & X & X & X & X & X\\

 & \textbf{External annotation tools}\textsuperscript{7} \tiny{(third-party tools)} & X & X\textsuperscript{8} & ~ & X & ~ & ~ &
~ & ~ & ~ & ~ & X & ~\\

 & \textbf{Exportation of temporalized data} & ~ & ~ & ~ & ~ & ~ & ~ &
X & X & ~ & ~ & ~ & ~ \\

 & \textbf{Internal annotations sharing} & ~ & ~ & ~ & ~ & ~ & ~ &
X & ~ & ~ & X & ~ & ~ \\

 & \textbf{External annotations sharing}\textsuperscript{9} & ~ & ~ & ~ & ~ & ~ & ~ &
X & X & ~ & ~ & ~ & ~ \\
\end{supertabular}

Notes:
\textbf{1}. Usually generated automatically with the possibility of
correction by the students. 
\textbf{2}. Translations into other languages are carried out in
collaboration with students (crowd sourced translation). 
\textbf{3}. Synchronized slides. 
\textbf{4}. Use of multiple choice embedded question. 
\textbf{5}. “Video Questions” is a new feature available in beta
version. There is also a possibility to choose the ending of a video. 
\textbf{6}. Navigation of the video through transcriptions. 
\textbf{7}. Usually VideoNot.es, those are the platforms featured on
its website. 
\textbf{8}. The use of the tool is promoted on the \underline{Coursera wiki page}.
\textbf{9}. In most cases, videos on YouTube can be used by the
external tools. 
\end{table*}

From our analysis, it appears first that features that facilitate the
comprehension of the discourse - such as the possibility to adjust the
video speed or activate subtitles and transcriptions - are largely
present in MOOCs. These tools seem to be important in a multicultural
context where subtitles or transcription are frequently produced in
collaboration with student in the case of translations to other
languages. Second, there is the use of interactive enrichments in MOOC
platforms, usually to have the video stop so that students answer a
question in order to verify the understanding of what had just been
explained. Third, if many MOOC platforms allow adding comments on
video lectures, annotations refering to a part of the video are only
possible via external tools. Fourth, the dedicated tools we analyzed
offer more features regarding the annotation process - such as
interactive timelines, export of annotated data, sharing of
annotation, etc. - than MOOC platforms.

Thus, most of the solutions presented in Table~\ref{table-annotations}
provide the possibility to develop pedagogical activities aiming to
achieve active reading with annotation from students. Indeed, this is
the main use of such tools by MOOC students, who are seeking a better
understanding of the proposed video material. Annotation-based active
reading can be carried out individually as well as collaboratively in
the majority of the tools. This last possibility is even more
significant: as students cannot always rely on having their
doubts/difficulties solved by the teacher, collaboration with peers
via annotation tools can increase their understanding along with
secondary benefits of developing cognitive capacities on learning from
video, observational skills and increasing focus and attention.

As far as our other classes of scenarios are concerned, performance
annotation was not observed in the MOOC context although it could lead
to an improvement of courses as it would be based on facilitated
self-reflection for the teacher or, even better, if made in
collaboration with students. Tasks involving annotation as assignment
were not observed either, though it is clear that the possibility of
critical exploration where students must find evidences to support
their thinking could be used within the MOOC context, even so with the
use of external annotation tools and peer evaluation (a well developed
practice already used regarding text assignments).

\section{\uppercase{Some challenges}}
\label{sec:challenges}

It appears that video material and its use in MOOCs are massive
nowadays; nevertheless a more advanced use of video enrichments and
more specifically of video annotations is still not a reality. From
the classes of scenarios we described earlier and the overview of the
current situation of e-learning and MOOC platforms, we would like to
put forward a number of challenges that we think should be addressed
in future versions of e-learning systems, linked with annotation
issues.

 \textbf{Manual production of annotations.} Manual
annotation processes raise specific ergonomic and usability issues,
all the more in the video domain where playing the document may
interfere with annotation entry. The variety of targeted devices like
mobile phones exacerbates these issues. Moreover, as mentioned above,
annotations are also an ideal vehicle for collaboration activities
around videos, in a synchronous~\cite{Nathan2008CollaboraTV:-Ma} or asynchronous
way. This brings some specifics issues, notably around the ergonomics
of video co-annotation; as well as privacy (e.g. the level of shared
information must be clearly displayed and tunable by users).

\textbf{Semi-automatic generation of annotations.} Most video-based
e-learning systems use only plain videos, sometimes fragmented into
small independent videos, providing only basic features. In order to
make these videos more accessible, e-learning platforms should
commonly provide features such as transcription or chaptering. Some
projects such as \underline{TransLectures} aim at providing automatic
or semi-automatic transcription of video, so that users may use the
transcription as entry points into the video, either for querying and
finding specific fragments, or as a simple navigation means.

 \textbf{Rich media and hypervideos.} Beyond the basic
video layout (side to side, overlay) that can be used to display the
video material, annotations can be used to enrich the video or to
produce whole new documents, such as
hypervideos~\cite{Chambel2006Hypervideo-and-} that are documents
combining video and assets originating from annotations. Challenges
here pertain to ergonomics, document modelling and (semi-)automatic
production, for instance through an annotation-guided summarization.

 \textbf{Video annotations related learning analytics.}
With MOOCs, learning analytics have become a major concern for all
organisations, by necessity - on this kind of scale, it is important
to take informed decisions - and by opportunity - we now have the
technological and processing capacity to capture and analyse the huge
amount of information generated by thousands of learners. The
annotation process and the annotations themselves offer an additional
source of information for learning analytics at a finer scale, that
could qualify as micro-analytics. Given the importance of video
resources, it is undoubtedly important to have precise feedback on its
reception. This new source of information could be used for example
in course re-engineering~\cite{Sadallah2013A-Framework-for}.

 \textbf{Annotation model and sharing.} Numerous tools
provide video annotation features, and many use custom data models for
storing annotation information
(\underline{Cinelab},
\underline{Exmeralda}…). However,
standardization efforts are underway to define more interoperable and
generic annotation models, able to encompass various annotation
practices on different source documents and to integrate well with the
current semantic web efforts
(\underline{OpenAnnotation}). 
Let us remark that some universities, mainly in the
US, are strongly committed to pushing forward and generalizing
annotation practices among students and faculty members, building
annotation ecosystems: Columbia~\cite{Bossewitch2011Teaching-and-Le},
Stanford~\cite{Pea2004The-Diver-proje} and
\underline{Harvard}.

 All these challenges share common concerns. First, mobile
phones and tablets have become important platforms for consulting
various resources, and among them, pedagogical resources. It is
important to propose the most complete experience on
annotation-enhanced e-learning platforms on all devices, and
especially on mobile ones, which have important constraints in terms
of display size and general capacity. Second, copyright and licensing
issues are even more stringent, since they concern not only the video
document (which has to be shareable to allow collaborative work), but
also the produced annotations. Clear licenses for this additional data
should be specified, hopefully with a bias towards openness and
reuse. Eventually, the question of accessibility - mainly for sensory
deficiencies - has to be considered as video annotations are clearly a
means to provide a better level of accessibility to video
content~\cite{Encelle2011Annotation-base}.

We have proposed four classes of scenarios illustrating how video
annotations can be used in e-learning contexts. To evaluate in what
measure these scenarios are feasible or already present, we have
reviewed a number of e-learning platforms (focusing on MOOCs) and
tools, in order to identify existing annotation features. It appears
that if some support already exists, there is still plenty of room to
efficiently implement the scenarios that go beyond simple active
reading, and a number of challenges related to video annotation still
remain. These challenges should be addressed in future versions of
e-learning systems, and we will tackle some of them in our future work
on the COCo platform\footnote{The authors of the paper are involved in
  the COCo project (\underline{Cominlabs Open Courseware}) based in
  University of Nantes, which is a recent initiative of the Cominlabs
  laboratory. The project goals are to build and animate a research
  platform for both disseminating and promoting rich media open
  courseware content.}.

\bibliographystyle{apalike}
{\small
\bibliography{csedu.bib}}
\vfill

\appendix
\onecolumn
\section{Webography}

You will find here the URLs referenced in the article, in alphabetical
order. Due to editing limitations, they could not be included as
hyperlinks in this version. The version of the article on the author's
website http://www.comin-ocw.org/ has them properly hyperlinked.\\
\\
Advene: http://www.advene.org/  \\
Annotated HTML: http://www.stanford.edu/group/ruralwest/cgi-bin/drupal/content/building-annotated-video-player-html  \\
Anvil: http://www.anvil-software.de/  \\
CLAS: http://isit.arts.ubc.ca/support/clas/  \\
Canvas Network: https://www.canvas.net/  \\
Cinelab: http://advene.org/cinelab/  \\
Cominlabs Open Courseware: http://comin-ocw.org/ \\
Coursera wiki page: https://share.coursera.org/wiki/index.php/Third-party\_Tools  \\
Coursera: https://www.coursera.org/  \\
EdX: https://www.edx.org/  \\
EliteSportsAnalysis: http://www.elitesportsanalysis.com/   \\
Exmeralda: http://www.exmeralda.org/  \\
Harvard: http://annotations.harvard.edu/  \\
Iversity: https://www.iversity.com/  \\
Khan Academy: https://www.khanacademy.org/  \\
Matterhorn Player: http://opencast.org/matterhorn/feature-tour/  \\
MediaNotes: http://www.cali.org/medianotes  \\
Mediathread: http://mediathread.ccnmtl.columbia.edu/  \\
MotionView Video: http://www.allsportsystems.com/  \\
Noldus: http://www.noldus.com/  \\
Open2Study: https://www.open2study.com/  \\
OpenAnnotation: http://www.w3.org/community/openannotation/  \\
Transana: http://www.transana.org/  \\
Translectures: http://www.translectures.eu/ \\
VideoANT: https://ant2.cehd.umn.edu/  \\
VideoNot.es: http://www.videonot.es/  \\
VideoNot.es: http://www.videonot.es/  \\
VideoTraces: http://depts.washington.edu/pettt/projects/videotraces.html  \\
VideoTraces: http://depts.washington.edu/pettt/projects/videotraces.html  \\
YouTube: http://www.youtube.com/  \\
Yovisto platform: http://www.yovisto.com/  \\

\end{document}